# Analysis of double-slit interference experiment at the atomic level

Jonathan F. Schonfeld

Harvard-Smithsonian Center for Astrophysics, Cambridge, Massachusetts 02138, USA

jschonfeld@cfa.harvard.edu

617-496-7604

ORCID ID# 0000-0002-8909-2401

**Abstract** I argue that the marquis characteristics of the quantum-mechanical double-slit experiment (point detection, random distribution, Born rule) can be explained using Schroedinger's equation alone, if one takes into account that, for any atom in a detector, there is a small but nonzero gap between its excitation energy and the excitation energies of all other relevant atoms in the detector (isolated-levels assumption). To illustrate the point I introduce a toy model of a detector. The form of the model follows common practice in quantum optics and cavity QED. Each detector atom can be resonantly excited by the incoming particle, and then emit a detection signature (e.g., bright flash of light) or dissipate its energy thermally. Different atoms have slightly different resonant energies per the isolated-levels assumption, and the projectile preferentially excites the atom with the closest energy match. The toy model permits one easily to estimate the probability that any atom is resonantly excited, and also that a detection signature is produced before being overtaken by thermal dissipation. The end-to-end detection probability is the product of these two probabilities, and is proportional to the absolute-square of the incoming wavefunction at the atom in question, i.e. the Born rule. I consider how closely a published neutron interference experiment conforms to the picture developed here; I show how this paper's analysis steers clear of creating a scenario with local hidden variables; I show how the analysis steers clear of the irreversibility implicit in the projection postulate; and I discuss possible experimental tests of this paper's ideas. Hopefully, this is a significant step toward realizing the program of solving the measurement problem within unitary quantum mechanics envisioned by Landsman, among others.

## 1. How does it happen?

Double-slit interference is a hallmark of quantum mechanics and has been observed repeatedly with projectiles as diverse as photons [1], electrons [2] and neutrons [3]. Related phenomena have been observed even with complex molecules [4]. The characteristic double-slit behavior, universally observed, is easy to describe. A single particle passes through a barrier with two open slits. On the downstream side of the barrier is a detector. The detector registers the arrival of the projectile at a single location. When this is repeated, the detector again registers the arrival of the projectile at a single location, but this time the location is different. When this is repeated again and again, it eventually becomes apparent that the detection positions are randomly distributed, and the distribution looks just like the familiar interference pattern in ripples on the surface of a water wave tank. More properly, the distribution is proportional to $|\psi|^2$, the absolute-square of the wavefunction of the particle after passing through the double slits (it is not *equal* to $|\psi|^2$ when normalized to the number of incoming particles; the proportionality constant reflects the detector's ability to react, and inefficiencies due to dissipation).

I can say with confidence that there is no consensus about how this happens. In the Copenhagen interpretation, the facts that a single particle induces a detection signature at only a single point ("wavefunction collapse"), that the points are randomly distributed, and that the distribution follows $|\psi|^2$ (Born rule) are taken as axiomatic. They are also taken as distinct from Schroedinger's equation, which governs how the wavefunction evolves between slits and detector. In the many worlds interpretation, all possible detections take place at the same time in different branches of reality, but it is still taken as axiomatic that within any one branch a single particle induces a detection signature at only a single point. Other interpretations attempt to attribute random-single-point-detection phenomenology to dynamical processes happening at layers of reality more fundamental than Schroedinger's equation.

This is a very frustrating situation, because Schroedinger's equation by itself accounts for an overwhelming variety of phenomena with spectacular accuracy. Why should it not be enough for explaining a few details about single-particle detection phenomenology? Have we overlooked something salient but universal about how detectors work? If we took this something into account, would it then become straightforward to explain the celebrated characteristics of the double slit experiment directly from Schroedinger's equation without interpretive axioms?

The purpose of this paper is to draw attention to one universal aspect of detector physics that has been overlooked in the quantum foundations literature, and to argue using a toy model that it can form the basis of an explanation of accepted quantum detection phenomenology entirely in terms of Schroedinger's equation alone. The overlooked aspect is that for any atom in a detector, there is a small but nonzero gap between its excitation energy and the excitation energies of all other relevant atoms in the detector (isolated-levels assumption). I readily acknowledge that this observation breaks no new ground in the fundamental description of detectors. It does, however, permit me to break new ground in the *analysis* of quantum detection. In particular, I am able to argue that the isolated-levels assumption, viewed through the lens of an elementary resonance analysis, accounts for the principal features of quantum detection that one typically explains using axioms: detection is observed to take place at a single point, the points are distributed randomly, and the random distribution follows the Born rule. Hopefully, this is a significant step toward realizing the program of solving the measurement problem within unitary quantum mechanics envisioned by Landsman [5], among others.

The present work is a paper on quantum foundations that draws on physical constructs and mathematical methods from quantum optics and cavity QED. In that respect the paper's techniques may be unfamiliar to those that care about its motivation, and its motivation may not be very urgent to those that are fluent in its techniques. I hope this doesn't deter either audience from approaching the paper. To this end, I have tried to make the exposition self-contained.

## 2. Model detector, part I: capture

The toy detector model that occupies the remainder of this paper is composed of atoms, each of which can be promoted to an excited state by an incoming particle, and then dissipate heat or decay into something like an escaping bright flash of light or some other strong signature. (The word "atom" is a convenience; it stands for anything localized, including molecules.) I posit that the single-atom decay/escape event is what counts as detection for the purpose of this paper. (In most real cases detection is heralded by behavior involving many atoms. I make the assumption that I can idealize this phenomenologically as a one-atom process because I will argue that the principal features of quantum detection that one typically uses axioms to account for are independent of the details of the detection signature.) I posit that dissipation or decay/escape timescales are long enough that we can speak of initial capture dynamics without reference to dissipation or decay/escape. I draw on the quantum optics literature [6] for the capture Hamiltonian,

$$\hbar^{-1}H = \sum_n \omega_n |n\rangle\langle n| + \sum_{\mathbf{k}} \left[ \omega_{\mathbf{k}} |\mathbf{k}\rangle\langle \mathbf{k}| + \frac{\varepsilon}{L^{3/2}} \sum_n \left( e^{i\mathbf{k}\cdot\mathbf{r}_n} |\mathbf{k}\rangle\langle n| + e^{-i\mathbf{k}\cdot\mathbf{r}_n} |n\rangle\langle \mathbf{k}| \right) \right], \quad (1)$$

where $\mathbf{k}$ is projectile wavevector, $n$ enumerates detector atoms, $|n\rangle$ describes a single excited atom, $\mathbf{r}_n$ is the position of atom $n$, $\varepsilon$ is a coupling constant, and $L$ is quantization volume. Different atoms have different resonant frequencies $\omega_n$ because local conditions (proximity of other atoms, Doppler shifts, etc.) vary. Section 8 illustrates this with the range of local conditions in a gaseous detector used in a neutron-interference experiment. In Eq. (1), all atoms interact with the incoming particle at the same time. (I have not made provision for a wavefunction's gradual advance through the detector medium, so I am assuming detection is dominated by the first atom to resonate. I am also ignoring the quantum-mechanical nature of the atomic positions $\mathbf{r}_n$. If one assumes detector atoms are much heavier than the particles to be detected, then treating their positions classically seems like a practical way to start. If the arguments that follow don't work even with that assumption, there's certainly no point going farther.)

To move ahead, I need to simplify: An incoming wavepacket takes the form

$$|\text{in}\rangle = \left(\frac{2\pi}{L}\right)^{3/2} \sum_{\mathbf{k}} \varphi(\mathbf{k}) |\mathbf{k}\rangle, \quad (2)$$

where momentum- and position-space wavefunctions $\varphi$ and $\psi$ are related by

$$\psi(\mathbf{r}) = \left(\frac{2\pi}{L^2}\right)^{3/2} \sum_{\mathbf{k}} e^{-i\mathbf{k}\cdot\mathbf{r}} \varphi(\mathbf{k}). \tag{3}$$

If |in> is normalized to unity then

$$\int |\psi(\mathbf{r})|^2 d^3r = \int |\varphi(\mathbf{k})|^2 d^3k = 1. \tag{4}$$

It follows that

$$\langle in|\hbar^{-1}H|n\rangle = \varepsilon\psi^*(\mathbf{r_n}). \tag{5}$$

So *if the Hamiltonian is restricted* to the space spanned by |in> and the {|n>}, and the incoming wavepacket is assumed narrowband, then one arrives at

$$\hbar^{-1}H = \omega_{in}|in\rangle\langle in| + \sum_n \omega_n |n\rangle\langle n| + \varepsilon \sum_n (\psi^*(\mathbf{r_n})|in\rangle\langle n| + \psi(\mathbf{r_n})|n\rangle\langle in|) \tag{6}$$

This is a very important restriction, as the next section will make plain. I will justify this restriction in Section 4.

## 3. Point detection and randomness

We can learn a lot from Equation (6). To make the notation even less unwieldy, recast it as the generic problem of a state space spanned by a single vector |p> and an orthogonal set {|n>} indexed by $n=1...$, with Hamiltonian

$$H = H_0 + H_I \tag{7}$$

where the uncoupled part is

$$\hbar^{-1}H_0 = \omega |p\rangle\langle p| + \sum \Omega_n |n\rangle\langle n| \tag{8}$$

and the coupling is

$$\hbar^{-1}H_I = |p\rangle\langle v| + |v\rangle\langle p|. \tag{9}$$

The vector |v> is a linear combination of the |n> and not necessarily normalized to unity. Because of the isolated-levels assumption, the frequencies {$\Omega_n$} are all different.

Now consider the elementary problem of what happens to a state that starts at |p> and then evolves according to H: To solve the problem, start with the eigenvalue condition

$$\lambda|e> = \hbar^{-1}H|e> = \hbar^{-1}H_0|e> + |p><v|e> + |v><p|e>. \tag{10}$$

Rearrange to form

$$|e> = (\lambda - \hbar^{-1}H_0)^{-1}|p><v|e> + (\lambda - \hbar^{-1}H_0)^{-1}|v><p|e>. \tag{11}$$

Project onto $<p|$ and onto $<v|$ to get simultaneous linear equations for $<v|e>$ and $<p|e>$. The determinant condition is a single equation for eigenvalue without reference to eigenvector,

$$\lambda - \omega = <v|(\lambda - \hbar^{-1}H_0)^{-1}|v> = \sum \frac{|v_n|^2}{\lambda - \Omega_n}. \tag{12}$$

The $v_n$ are the components of $|v>$ with respect to the basis $\{|n>\}$ (and don't forget $H_0$ is diagonal in the orthogonal basis $\{|p>, |n>\}$). The normalized eigenvector associated with eigenvalue $\lambda$ is then

$$|\lambda> = \left[|p> + \sum \frac{v_n}{\lambda - \Omega_n}|n>\right]\left[1 + \sum \frac{|v_n|^2}{(\lambda - \Omega_n)^2}\right]^{-\frac{1}{2}}. \tag{13}$$

Finally, a state that starts at $|p>$ becomes

$$|p(t)> = \sum_\lambda e^{i\lambda t}\left[|p> + \sum_n \frac{v_n}{\lambda - \Omega_n}|n>\right]\left[1 + \sum_n \frac{|v_n|^2}{(\lambda - \Omega_n)^2}\right]^{-1} \tag{14}$$

at time $t$.

This general solution is important because it leads to the following conclusion: If the interaction is weak and the projectile is not resonant with any atom, i.e. if $\omega$ is not close to any $\Omega_n$, then, according to Eq. (12), every $\lambda$ is equal to $\omega$ or some $\Omega_n$ with remainder $O(v^2)$, and therefore

$$|p(t)> = e^{i\omega t}|p> + O(v). \tag{15}$$

Since the remainder is small, it is safe to say that essentially all observers will agree that nothing happens to the projectile, whether or not one cares to identify wavefunction square norm with probability. (This may sound like assuming a probability interpretation of wavefunction square norm a priori when that's really what I want to derive. I will try to justify the neglect of small remainders without an a priori probability interpretation in Section 5.) However, if the "projectile" is resonant with some atom (call it $m$), i.e. if $\omega$ is close ($<<O(v_m)$) to $\Omega_m$, then, according to Eq. (12), every $\lambda$ but two is equal to some $\Omega_n$ with remainder $O(v^2)$. The two exceptions are

$$\frac{1}{2}(\omega + \Omega_m) \pm |v_m| + O(v^2), \tag{16}$$

so that

$$|p(t)\rangle = e^{it(\omega+\Omega_m)/2}\{|p\rangle \cos(|v_m|t) + i|m\rangle \sin(|v_m|t)\} + O(v). \qquad (17)$$

Again, since the remainder is small, it is safe to say (again see Section 5) that essentially all observers agree that |p> is locked into resonant oscillation with |m>, and all other atoms are largely irrelevant. We assume the frequencies $\{\Omega_n\}$ and $\omega$ are randomly distributed and uncorrelated due to inevitable inaccuracies in setting up the experiment. It follows that the excited atom's index $m$ is also a random variable.

(Mathematically, the equations in this section also describe the multi-mode weakly coupled Purcell effect [7], [8]. This raises the interesting possibility that processes in this paper could be emulated by a suitable cavity-QED experiment, where cavity modes serve as proxies for detector atoms.)

In the notation of Section 2, the condition "close to resonance" amounts to $|\omega_{in}-\omega_n|\ll\varepsilon|\psi(\mathbf{r}_n)|$. So the probability of achieving resonance is

$$P_{\text{resonance}} = 2\varepsilon|\psi(\mathbf{r}_n)|/\Omega, \qquad (18)$$

where $\hbar\Omega$ is the range of energies in a statistical ensemble of projectiles. (I am assuming the *intrinsic* energy width of a projectile wavepacket is $\ll \hbar\Omega$, in fact small enough to not matter for any of the calculations that follow. A detailed accounting of how the physical projectile wavepacket itself forms is beyond the scope of this paper.) The fact that this scales with the absolute value of the projectile wavefunction suggests this model is on the right track, but it's not quadratic as expected from the Born rule. The rest of the Born rule will emerge from the work of Sections 6 and 7.

## 4. Rationale for restricting the capture Hamiltonian to the span of |in> and {|n>}

Let us pause briefly to address the issue raised at the very end of Section 2. To arrive at Equation (6), I ignored the possibility that an excited atom can decay back into a projectile state other than |in>. I can do that because, following the results of the preceding section, only states resonant with |in> matter for detection. Presumably an incoming projectile state other than |in> would mean a different microscopic state of the atoms at the double slit that shaped |in> in the first place, and therefore an overall energy not resonant with $\omega_{in}$.

## 5. Rationale for neglecting small quantum-state remainders

Let us further pause briefly to address the assumption that we can neglect small state remainders, as in Section 3, without implicitly endowing square-norm with an a priori probability interpretation. I can only offer a "philosophical" argument: One must consider that an observer is himself made of wavefunctions. Under Hamiltonian evolution (Schroedinger's equation) wave functions can spread out into space or coalesce into bound dynamic entities.

Those coalesced entities can be exceedingly complex, to the point of stumbling upon their own versions of thinking, acting and measuring, i.e. becoming us. The only fundamental metric available for distinguishing one such being from another, or from its surroundings, is wavefunction square norm. "Things" are concentrations of square norm. A big square norm should somehow translate into a coalesced entity's experience. It is hard to say how this works for square norms anywhere on a spectrum between 0 and 1, but fur square norm very close to 0 or very close to 1 it should be simple: approximately 0 means nothing happens, approximately 1 means something always happens. I have put quotes around "philosophical" because in principle you could run a huge computer simulation and see for yourself the simulated wavefunctions coalescing, then learning how to think and measure, and finally measuring.

## 6. Model detector, part II: emission

Let us now focus on what happens after capture. I start with a simplification for convenience. If the separation between $\omega_{in}$ and a particular $\omega_n$ is much smaller than the mean level spacing, then, following Section 3, I can reduce the capture Hamiltonian in Equation (6) to an effective two-state operator

$$\begin{pmatrix} \omega_{in} & \varepsilon\psi^*(\mathbf{r_n}) \\ \varepsilon\psi(\mathbf{r_n}) & \omega_n \end{pmatrix}. \tag{19}$$

Now I incorporate emission of a detection signature, by modifying (19) phenomenologically to include a decay width $\Gamma$,

$$\begin{pmatrix} \omega_{in} & \varepsilon\psi^*(\mathbf{r_n}) \\ \varepsilon\psi(\mathbf{r_n}) & \omega_n + i\hbar^{-1}\Gamma \end{pmatrix}. \tag{20}$$

For resonant projectile and large $\Gamma$, i.e. for

$$|\omega_{in} - \omega_n| \ll \varepsilon|\psi(\mathbf{r_n})| \ll \hbar^{-1}\Gamma, \tag{21}$$

the two eigenvalues of Equation (20) reduce to

$$i\hbar^{-1}\Gamma, \quad i\varepsilon^2|\psi(\mathbf{r_n})|^2/\hbar^{-1}\Gamma \tag{22}$$

(removing a common $\omega_{in}$). The corresponding eigenvectors are very close to (0,1) and (1,0), respectively. Since (1,0) is the initial condition in this projectile scenario, one can conclude that nearly 100% of the projectile very slowly leaks into the state of emission of a detection signature. Competition between this slow leakage and thermal dissipation (next section) will account for the missing power of $|\psi(\mathbf{r}_n)|$ needed to turn Equation (18) into the Born rule.

## 7. Model detector, part III: dissipation

Alongside emission, the resonant atom experiences thermal dissipation due to random perturbations from its surroundings. Phenomenologically, it seems like a reasonable ansatz to model the effect of these perturbations on the atom as a Markov random walk through a hypothetical fine structure of nearby internal energy levels. (I do not provide here a unitary account of the dissipation mechanism, because I am arguing that the principal features of quantum detection that one typically uses axioms to account for are independent of the details of the dissipation mechanism.) From the general theory of random walk [9] – and without any need to appeal to an a priori relation between quantum amplitudes and probabilities – it follows that the statistics of energy-level occupation evolve in time according to a diffusion law. In particular, after time $\tau$ the probability distribution of this random walk in $\omega$ must take the general form

$$P(\omega - \omega_n, \tau) = \frac{1}{\sqrt{2\pi g \tau}} \exp - \frac{(\omega - \omega_n)^2}{2\pi g \tau} \qquad (23)$$

for some phenomenological parameter $g$.

As a final phenomenological ansatz, I suppose that $|\omega-\omega_n|$ must be less than some rough threshold $G/2$ for the process described by the second eigenvalue in Eq. (22) to proceed. The probability of satisfying this inequality is roughly

$$GP(0, \tau) = \frac{G}{\sqrt{2\pi g \tau}}. \qquad (24)$$

In order for detection to take place, this inequality must hold until a time equal to one over the second eigenvalue in Eq. (22). The probability of that happening is roughly

$$\frac{G}{\sqrt{2\pi g (\hbar^{-1}\Gamma/\varepsilon^2 |\psi(\mathbf{r}_n)|^2)}} = \frac{G\varepsilon|\psi(\mathbf{r}_n)|}{\sqrt{2\pi g \hbar^{-1}\Gamma}} \approx \varepsilon|\psi(\mathbf{r}_n)|\sqrt{\frac{\hbar^{-1}\Gamma}{2\pi g}} \qquad (25)$$

I have taken the liberty of replacing $G$ by $\hbar^{-1}\Gamma$ (it seems intuitive that, at the very least, $G=O(\hbar^{-1}\Gamma)$).

The total detection probability for atom $n$ is the product of the right-hand-sides of Equations (18) and (25), i.e.

$$P_{\text{detection}} = \varepsilon^2 |\psi(\mathbf{r}_n)|^2 \sqrt{\frac{2\hbar^{-1}\Gamma}{\pi g \Omega^2}} = |\psi(\mathbf{r}_n)|^2 \left(\frac{\varepsilon^2}{\omega_{\text{in}}^2}\right)\left(\omega_{\text{in}}^2 \sqrt{\frac{2\hbar^{-1}\Gamma}{\pi g \Omega^2}}\right). \qquad (26)$$

The far-right expression in Eq. (26) is factored into $|\psi|^2$, times an interaction volume that depends only on the incoming capture process, times a factor independent of the capture process that can be interpreted as quantum efficiency. In this form one recognizes Equation (26) as the Born rule, where, as usual, the proportionality constant multiplying $|\psi|^2$ reflects the detector's ability to react, and inefficiencies due to dissipation. Presumably, a more careful analysis of the phenomenological

parameters in Equation (26) would recast the multiplier of $|\psi|^2$ as something even more familiar, but this is beyond the scope of this paper.

## 8. Neutron interference example

Let us assess how closely a neutron interference experiment conforms to the analysis in this paper. Neutron interference may seem like an odd choice at this point because the Hamiltonian in Equation (1) is obviously better suited for describing photon detection. However, in a well-known experiment [3], very slow neutrons are detected by $B^{10}F_3$ gas at standard temperature and pressure, and an ideal gas is very easy to analyze.

The detection mechanism is

$$B^{10} + n \rightarrow (B^{11})^* \rightarrow Li^7 + \alpha. \qquad (27)$$

A zero-energy resonance is inserted to account for an anomalously large reaction cross section. The excitation energy of the presumed intermediate state is

$$\Delta((B^{11})^*) = [m(B^{10}) + m(n) - m(B^{11})]c^2 = 12 \text{ MeV}. \qquad (28)$$

Other basic experimental parameters are
- Reaction cross section $\sigma \sim 4000$ bn $= 4 \times 10^{-25}$ m$^2$ [10].
- Incoming wavenumber $k = 2\pi/(20\text{Å}) = \pi \times 10^9$ m$^{-1}$.
- Experimental wavenumber spread $\delta k = (2\times 0.7\text{Å}/20\text{Å}) \times k = 7\% \, k$.
- Detector molecular number density $\rho \sim (3 \times 10^{-9} \text{ m})^{-3}$ [11].
- Typical component of BF$_3$ velocity along projectile direction $v = [k_B T/m(BF_3)]^{-1/2} \sim 300$ m/s.
- Width of projectile wavefunction transverse to direction of motion at detector $w \sim 400$ $\mu$m = 0.4 mm.

These values imply the following:
- Total volume occupied by the incoming wavepacket $\sim w^2(2\pi/\Delta k)$, where $\Delta k$ is the intrinsic width of the neutron wavepacket in k-space. Taking $\Delta k = \delta k$ just for numerical definiteness, this means total occupied volume $\sim 4 \times 10^{-15}$ m$^3$.
- This determines the typical wavefunction magnitude 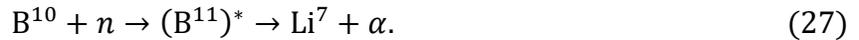 $|\psi| \sim (4 \times 10^{-15} \text{ m}^3)^{-1/2} \sim 2 \times 10^7$ m$^{-3/2}$.
- This volume also contains $\sim \rho \times (4 \times 10^{-15} \text{ m}^3) \sim 2 \times 10^{11}$ BF$_3$ molecules.
- The Doppler-driven spread in $(B^{11})^*$ excitation energy across this sample $S \sim (v/c) \times 12$ MeV $\sim 12$ eV. (This is a lower bound because other effects could contribute to $S$.)
- So the typical separation between adjacent excitation levels $\delta \sim (12 \text{ eV})/(2 \times 10^{11}) = 6 \times 10^{-11}$ eV.
- The neutron kinetic energy $E = k^2 \hbar^2/2m \sim 0.2$ meV.
- Neutron energy spread $\hbar \Omega \sim 2E \delta k/k \sim 3 \times 10^{-5}$ eV.

The upshot of all this is to confirm the assumption from preceding sections that $\delta \ll \hbar\Omega \ll S$. Since $\delta$ is proportional to $\Delta k$, and $\Delta k \ll \delta k$, the assumption has extra margin.

To confirm the consistency of the close-to-resonance condition (first inequality in Eq. (21)), we need to try to estimate $\varepsilon$ from the nuclear data. This is not straightforward because the capture Hamiltonian (1) is so electromagnetic-like in its basic structure. I try to estimate $\varepsilon$ from $\sigma$ by dimensional analysis. There is no unique way to do this, so look at two alternative approaches. The first approach assumes the cross section arises from conventional perturbation theory in $\varepsilon$. The only cross-section-dimension combination of neutron mass, wavenumber and Planck's constant also quadratic in $\varepsilon$ is $\varepsilon^2 m^2/k^3\hbar^2$, implying $\varepsilon \sim 3 \times 10^{-6}$ s$^{-1}$m$^{3/2}$. The second approach assumes that the large neutron-capture cross section indicates a near-zero-energy resonance and therefore approximately scales like $k^{-2}$ [12]. But this power law already has dimensions of cross section, so this second approach cannot determine $\varepsilon$. Thus we only have the first approach, which implies $\hbar\varepsilon|\psi| \sim 4 \times 10^{-14}$ eV $\ll \delta$. This supports the notion that the incoming particle can distinguish between different detector molecules by resonant frequency.

## 9. Hidden variables critique

It seems clear from the foregoing that if atomic positions and projectile parameters could be pre-specified with sufficient precision, then detection could be made to take place at the same location, trial after trial, regardless of how $|\psi|$ depends on position. As a practical matter this is very difficult to arrange, because detector atoms are constantly in thermal motion. But as a theoretical matter it suggests a deterministic local hidden variables scenario, in which atomic positions and projectile parameters are the hidden variables. Narrowly speaking, this is not a problem because so far I have only discussed single-particle detection, not the two-particle cases that drive local hidden variables to well-known contradictions [13]. However, that doesn't make the underlying concern go away, so I now extend the foregoing to Einstein-Podolsky-Rosen (EPR) type experiments and show directly that the same hidden variables in that case do not act locally.

In the EPR scenario, an unstable spin-zero source decays into two particles ("near" and "far") that exit in opposite directions. Total spin zero implies that the outgoing state entangles the spins of the decay products in the usual way, $\{|up\rangle|down\rangle+|down\rangle|up\rangle\}/2^{1/2}$. According to the EPR paradox, if a measurement of the near spin produces the outcome up, then a subsequent measurement of the far spin must produce the outcome down. If the direction of the magnet that defines "up" and "down" in the near Stern-Gerlach apparatus is rotated, the measured value of the far spin must rotate accordingly.

To relate this to the model in this paper, consider that a Stern-Gerlach apparatus is basically a re-purposed position detector: If, say, an incoming particle is detected on the right half of the near Stern-Gerlach apparatus, it must have been deflected that way by the apparatus magnetic field, and therefore the particle spin in resonance must have pointed up. In the same way, if $\{|up\rangle|down\rangle+|down\rangle|up\rangle\}/2^{1/2}$ approaches the near Stern-Gerlach detector, it will resonate with only one detector atom, say $|m\rangle$, which is on, say, the right. The result of that resonant capture must then be $|m\rangle|down\rangle$ (i.e., resonance with $|m\rangle$ is "winner take all"), and when detection is subsequently observed at the far apparatus, it can only take place on the down side. Functionally, this is exactly the same as the canonical EPR outcome: The second measurement depends on the setup and outcome of the first. (Even though this argument requires saying that one detection comes before the other, and "before" is not a relativistically invariant concept, the final conclusion can be stated in relativistically invariant terms: spins detected in the two detectors must be random but opposite.) A similar invocation of winner-take-all resonance suffices to show compatibility with more sophisticated variants of EPR (for example, [14]).

## 10. Projection postulate critique

Von Neumann's formalism [15] asserts that quantum-mechanical measurement requires application of a projection operator. Since projection operators can't be inverted, measurement in the Von Neumann framework must be inherently irreversible.

According to the analysis in this paper, measurements are *not* equivalent in every way to projections. Projection is obviously an excellent effective theory; one probably has to do a somewhat contrived experiment (see next section) to realize a counterexample. But it is wrong to take the effective theory at face value "all the way down." In other words, irreversibility of measurement is an idealization.

The key mathematical step here is the state restriction that leads to the reduced capture Hamiltonian (Equation (6)). The transition from |in> to the excited state |m> is invertible because the energetics at the double slit guarantees that anything, other than |in>, that |m> might revert to has a slightly different energy and therefore can't resonate with |m> (Section 4). A slightly different wavepacket exiting from the double slit would resonate with a different detector atom (or no atom), so what happens to a generic wavepacket can't involve a projection – in fact can't be irreversible - because this process manifestly does *not* map many states to one.

## 11. Experimental tests and generalization to other detector types

To summarize, I argue that the marquis characteristics of the double-slit experiment can be derived from Schroedinger's equation by taking into account the isolated-levels nature of real detectors. Here are possible experiments to probe this picture:

- As discussed above, if atomic positions and projectile parameters could be pre-specified with sufficient precision, then detection could be made to take place at the same location, trial after trial, regardless of how $|\psi|$ depends on position.
- If dissipation could be suitably suppressed, then, following Eq. (18), detection probability could be made to be proportional to $|\psi|$ rather than $|\psi|^2$.
- If the incoming particle's energy could be defined narrowly enough, and the spread of atomic frequencies $\{\omega_n\}$ could be biased enough as a function of position, then detection probability could be made to vanish within a spatially extended region, regardless of how $|\psi|$ depends on position.
- If the detector medium could be made sufficiently rarified, so that $\Omega <$ typical spacing between resonance energies, one could observe a crossover where the detector simply runs out of the statistics that drive the Born rule.
- Suppose the incoming wavefunction has transverse length scale $L$. Then, for fixed wavefunction depth (determined by bandwidth) the mean spacing between detector atom energy levels scales like $L^{-2}$ but the mean wavefunction magnitude $|\psi|$ scales like the square root $L^{-1}$. So there could be an observable crossover between values of *L* for which the left-hand inequality in Eq. (21) could be satisfied by only one atom and values for which the inequality could typically be satisfied for more than one atom.

It may be disconcerting to read that the results of a quantum detection experiment can be repeatable trial after trial. It seems to go against all that we've learned about quantum mechanics. But all that we know about quantum detection is based on experiments in which the atoms in detector, slit apparatus and projectile source are *not* held in the same positions and states, trial after trial. If positions and states could be well-controlled, then detection results *would* come out the same trial after trial, and that would in no way contradict what experiment has taught us up to this point.

The model in this paper seems specific to detectors that exploit capture-emit mechanisms. Do the general conclusions in this paper apply also to detectors that do not? Based on the following example, I conjecture that they do apply. Consider a cloud chamber. This is not a capture-emit device; instead, an incoming particle ionizes a succession of vapor molecules to produce a track. Mott's wave-mechanics analysis [16] explains why cloud-chamber detections resolve into tracks but doesn't explain how a track starts in the first place. In this case the first ionization is still characterized by an effective Hamiltonian similar to Equation (1), but now each excited state describes an exiting triple of free projectile, free ion and free electron. Instinctively, we think of such states as forming a continuum, but in fact the admissible momenta of exiting projectile, ion and electron are quantized because the detector must be in a finite box. Thus only discrete combinations of individual projectile, ion and electron momenta can conserve overall momentum; and those combinations are prime suspects for the discrete states |n>. I conjecture that such discreteness – either explicit for the model in this paper, or implicit for the cloud chamber – can always be identified in any practical single-quantum detector, and is what universally drives the probabilistic phenomenology of single-projectile detection in unitary quantum mechanics, independent of interpretation.

*Acknowledgment: I am grateful for helpful feedback from James Babb, Peter Drummond and the referees.*

**References**

1. Aspden, R., Padgett, M., Spalding, G.: Am. J. Phys. 84, 671 (2016)
2. Tonomura, A. et al: Am. J. Phys. 57, 117 (1989)
3. Zeilinger, A. et al: Single- and double-slit diffraction of neutrons. Rev. Mod. Phys. 60, 1067 (1988)
4. Hackermuller, L., et al: Nature 427, 711 (2004); Gerlich, S. et al: Nature Communications 2, 263 (2011)
5. Landsman, K.: Foundations of Quantum Theory. Springer, (2017)
6. Kyrola, E., Eberly, J.: Quasicontinuum effects in molecular excitation. J. Chem. Phys. 82, 1841 (1985)
7. Purcell, E.: Spontaneous emission probabilities at radio frequencies. Phys. Rev. 69, 681 (1945)

8. Sundaresan, N. et al: Beyond strong coupling in a multimode cavity. Phys. Rev. X5, 021035 (2015)
9. Chandrasekhar, S.: Stochastic Problems in Physics and Astronomy. Rev. Mod. Phys. 15, 1 (1943)
10. Sears, V.: Neutron scattering lengths and cross sections. Neutron News 3, 26 (1992)
11. Honeywell Corporation: Honeywell Boron Trifluoride Technical Information.
12. Taylor, J.: Scattering Theory. John Wiley & Sons, New York (1972)
13. Bell, J.: On the Einstein-Podolsky-Rosen paradox. Physics, 1(3) 195 (1964)
14. Franson, J.: Bell Inequality for Position and Time. Phys. Rev. Lett. 62, 2205 (1989)
15. Von Neumann, J.: Mathematical Foundations of Quantum Mechanics. Princeton University Press, Princeton (1955)
16. Mott, N.: The Wave Mechanics of $\alpha$-Ray Tracks. Proc. R. Soc. Lond. A, 126, 79 (1929)